\documentclass[twocolumn,showpacs,preprintnumbers,amsmath,amssymb,12 pt.,double-spaced]{revtex4}

\usepackage{graphicx}
\usepackage{dcolumn}
\usepackage{bm}
\usepackage{color}

\begin{document}

\preprint{APS/123-QED}

\title{Suppression of backward scattering of Dirac fermions in iron pnictides Ba(Fe$_{1-x}$Ru$_x$As)$_2$}

\author{Y. Tanabe$^1$}
 \thanks{Corresponding author: youichi@sspns.phys.tohoku.ac.jp}

\author{K. K. Huynh$^2$}

\author{T. Urata$^2$}

\author{S. Heguri$^2$}

\author{G. Mu$^2$}

\author{J. T. Xu$^1$}

\author{R. Nouchi$^1$}

\author{K. Tanigaki$^{1, 2}$}
\thanks{Corresponding author: tanigaki@sspns.phys.tohoku.ac.jp}

\affiliation{$^1$WPI-Advanced Institutes of Materials Research, Tohoku University, Aoba, Aramaki, Aoba-ku, Sendai, 980-8578, Japan}

\affiliation{$^2$Department of Physics, Graduate School of Science, Tohoku University, Aoba, Aramaki, Aoba-ku, Sendai, 980-8578, Japan}

\date{\today}

\begin{abstract}
We report electronic transport of Dirac cones when Fe is replaced by Ru, which has an isoelectronic electron configuration to Fe, using single crystals of Ba(Fe$_{1-x}$Ru$_x$As)$_2$.
The electronic transport of parabolic bands is shown to be suppressed by scattering due to the crystal lattice distortion and the impurity effect of Ru, while that of the Dirac cone is not significantly reduced due to the intrinsic character of Dirac cones.  
It is clearly shown from magnetoresistance and Hall coefficient measurements that the inverse of average mobility, proportional to cyclotron effective mass, develops as the square root of the carrier number (n) of the Dirac cones.
This is the unique character of the Dirac cone linear dispersion relationship.
Scattering of Ru on the Dirac cones is discussed in terms of the estimated mean free path using experimental parameters.
\end{abstract}

\maketitle

\section{Introduction}
Transport phenomena of massless fermions are intriguing subjects in the physics of Dirac cones and have become a very important frontier research area. 
A Dirac cone characterized by a degeneracy point in the linear band dispersion is known to be realized in materials with special geometrical symmetries, such as the K-K' degenerate points in graphene \cite{Castro}, topological insulators (TIs) with the Z$_{2}$ symmetrical constraint \cite{Kane}, some organic conductors \cite{Tajima} with a degenerate crossing band, and iron pnictides with different d$_{xy}$-d$_{xz}$ orbital symmetry \cite{Fukuyama, Ran, Morinari, Richard, Harison, Shimojima, Ming, Kim, Khuong, Imai, Bhoi, Pallecchi, YZhang}.
One of the significant features of electronic transport in the Dirac cone states is the large suppression of backward scattering as a consequence of the Berry's phase.

In iron pnictides, the Dirac cone is created in pairs as the node of the spin density wave (SDW) ordering states caused by the orbital symmetry in the Fe-multi band system \cite{Ran}.
The resulting Dirac cones have pseudo-spin chirality caused by the chirality compensation between the Dirac cones and the parabolic hole pockets \cite{Morinari}.
These Dirac cones might therefore cause a large suppression in backward scattering.
The situation can be compared to that observed for graphene, where the unit cell has two sublattices generating the Dirac cones in pairs at points K and K$^{\rm \prime}$ in the first Brillouin Zone with opposite chirarities \cite{Castro}.
The absence of backward scattering holds for the geometrically identical K and K$^{\rm \prime}$ points, but it is broken in the case of intervalley scattering due to the opposite chirarity \cite{Ando}.
In recent studies on topological insulators, a single Dirac cone emerges to create a helical spin texture at the time reversal invariant momentum due to spin orbital coupling, and therefore the locked spin angular momentum prohibits backward scattering \cite{Kane}.

In iron pnictides, almost no superconducting fluctuation has been reported in the underdoped regime of iron pnictide superconductors\cite{Matusiak}, which is in strong contrast with what has been encountered for the high temperature superconducting cuprates\cite{Ong1, Ong2, Corson, Armitage, Nakamura}. 
If the high mobility carrier could exist behind the Cooper paired electronic states, the phase coherence of Cooper pairs would be much enhanced, which can result in a robust superconducting state against thermal fluctuations.
This is significantly important in the physics of the high temperature superconducting materials. 

Recently, a pronounced contribution of parabolic bands to electronic transport has been reported in Shubnikov-de Haas oscillation and Hall coefficient measurements for high quality single crystals made by annealing under BaAs\cite{Terashima, Ishida}.
Recent magnetoresistance measurements have reported that the Dirac cone persists in the underdoped regime of Ba(Fe$_{1-x}$Ru$_x$As)$_2$ \cite{Tanabe}.
Considering the present situation established in the Dirac cone states in  Ba(Fe$_{1-x}$Ru$_x$As)$_2$, it is very important to clarify the scattering effect of the Dirac cone states of this family from the viewpoint of electric transport in detail, although the contribution of the parabolic bands must be suppressed to extract the intrinsically unique carrier transport of the Dirac cone in a complex multi band system of iron pnictide materials.
In this research, we have studied this intriguing electronic state in the Fe-multi band system using Ru-substituted Ba(Fe$_{1-x}$Ru$_x$As)$_2$.
For this purpose, single crystals of Ba(Fe$_{1-x}$Ru$_x$As)$_2$ were rapidly quenched at around 1050 - 1100$^{\circ}$C, which can introduce a significant lattice distortion effect as discussed in detail in the text.
It will be shown that the mobility is greatly reduced by scattering due to crystal lattice distortion in the case of the parabolic bands, while that of the Dirac cone is not significantly influenced as the loss of backward scattering.
We will show that the inverse of the average mobility (1/$\mu$$_{\rm ave}$), proportional to the cyclotron effective mass ($m^{\ast}$), evolves as a function of square root of the carrier number, $n$, of the Dirac cone.
The ${\mathstrut n}$ dependence of $m^{\ast}$ is the unique character of the massless Dirac quasiparticles.
It will be shown that the estimated mean free path is constant for the whole region of $x$ and is not correlated with the Ru-Ru distance.

\section{Experimental}
Single crystals of Ba(Fe$_{1-x}$Ru$_x$As)$_2$ were grown by a flux method using FeAs \cite{Canfield, Tanabe}.
The quality of the single crystals was studied by synchrotron X-ray diffraction measurements at the beam line BL02B2, SPring-8.
Electric resistivity $\rho$ measurements were also carried out using a four-probe method to check the quality of the samples.
The Ru concentration $x$ for Ba(Fe$_{1-x}$Ru$_x$As)$_2$ was determined by employing the relationship between the c-axis lattice constant and $x$ as reported in Ba(Fe$_{1-x}$Ru$_x$As)$_2$ \cite{Canfield, Tanabe}.
The dependence of in-plane transverse magnetoresistance ($R_{\rm M}$) and the Hall coefficient ($R_{\rm H}$) on magnetic field $B$ was measured using a four-probe method in $B$ of -9 - 9 T at various temperatures (T) between 2 and 300 K.

\section{Results and discussions}
\begin{figure}
\begin{center}
\includegraphics[width=0.9\linewidth]{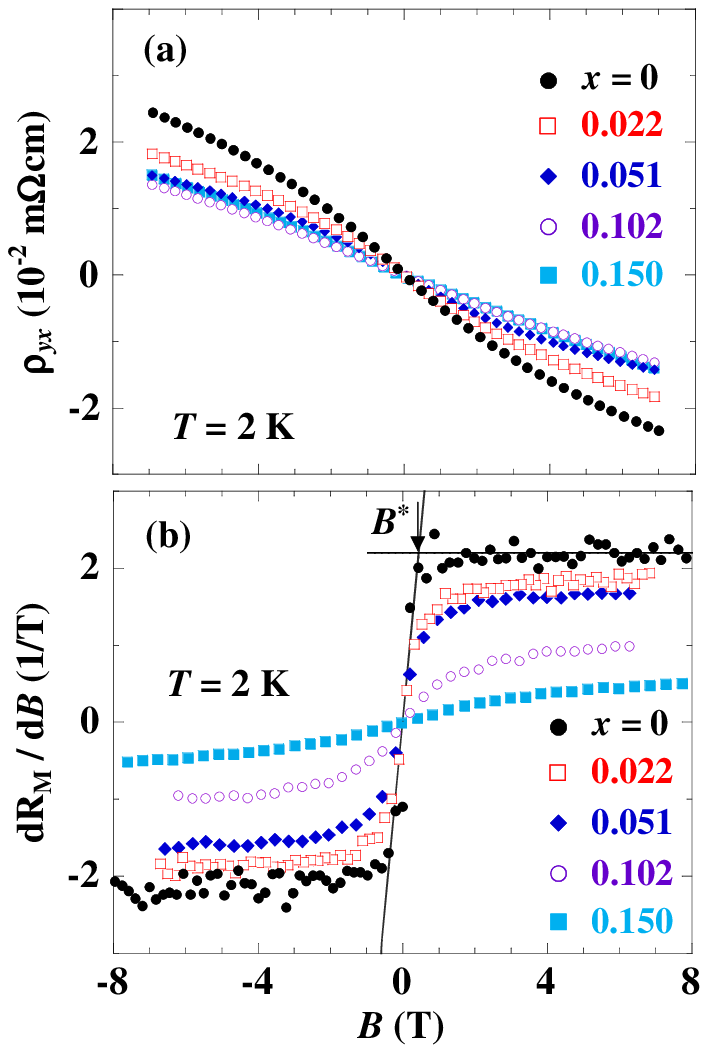}
\caption{(color online) Magnetic field (B) dependence of the Hall resistivity ($\rho$$_{yx}$) (a) and the derivative of magnetoresistance (d$R_{\rm M}$/d$B$) (b) for Ba(Fe$_{1-x}$Ru$_x$As)$_2$ with $x$ between 0 and 0.150 at 2 K.}
\end{center}
\end{figure}

Figures 1 (a) and (b) show the $B$ dependencies of Hall resistivity ($\rho$$_{yx}$) and the derivatives of magnetoresistances (d$R_{\rm M}$/d$B$) for Ba(Fe$_{1-x}$Ru$_x$As)$_2$ with $x$ between 0 and 0.150 at 2 K.
For $x$ = 0, $\rho$$_{yx}$ developed linearly with $B$ in the low $B$ regime, but showed deviation from the anticipated straight line.
This is consistent with the nonlinear behavior of $\rho$$_{yx}$ in the multiband nature of Ba(FeAs)$_2$ \cite{Shen, Kuo}.
The nonlinear behavior of $\rho$$_{yx}$ with $B$ was suppressed by Ru doping and an almost linear $\rho$$_{yx}$ was observed at $x$ = 0.150.
The value of d$R_{\rm M}$/d$B$ developed linearly with $B$ but was saturated above 2 T for $x$ = 0, which indicates that the $B$ dependence of $R_{\rm M}$ from $R_{\rm M}$ $\propto$ $B^2$ to $R_{\rm M}$ $\propto$ $B$ had changed.\cite{Khuong}.
The crossover magnetic field ($B^{\ast}$) between $R_{\rm M}$ $\propto$ $B^2$ and $R_{\rm M}$ $\propto$ $B$ increased with an increase in $x$, while its absolute value of d$R_{\rm M}$/d$B$ decreased with an increase in $x$ \cite{Tanabe}.

In order to clarify the contribution of Dirac fermions to the electronic transport, we analyzed the results of magnetoresistance and Hall coefficient using a two-carrier-type semiclassical approximation.
In the low $B$ limit, the zero-field resistivities ($\rho$(0), $R_{\rm M}$ and $\rho$$_{yx}$) are described as
\begin{eqnarray}
 \rho(0) &=& \frac{1}{e(n_e\mu_e + n_h\mu_h)} \label{1},\\
 R_{\rm M} &=& \frac{\rho(B) - \rho(0)}{\rho(0)} = \frac{n_en_h\mu_e\mu_h(\mu_e + \mu_h)^2B^2}{(n_e\mu_e + n_h\mu_h)^2} \label{2},\\
 \rho_{yx} &=& \frac{(-n_e\mu_e^2 + n_h\mu_h^2)B}{e(n_e\mu_e + n_h\mu_h)^2} \label{3},
\end{eqnarray}
where $n_{\rm e}$ and $n_{\rm h}$ are the carrier numbers, and $\mu$$_e$ and $\mu$$_h$ are the mobilities of electrons and holes, respectively.
The resulting fact that $R_{\rm M}$ and $\rho$$_{yx}$ are quadratic and linear against $B$ in the low $B$ regime is consistent with the two-carrier-type semiclassical approximation under a low $B$ strength.

\begin{figure*}
\includegraphics[width=1.0\linewidth]{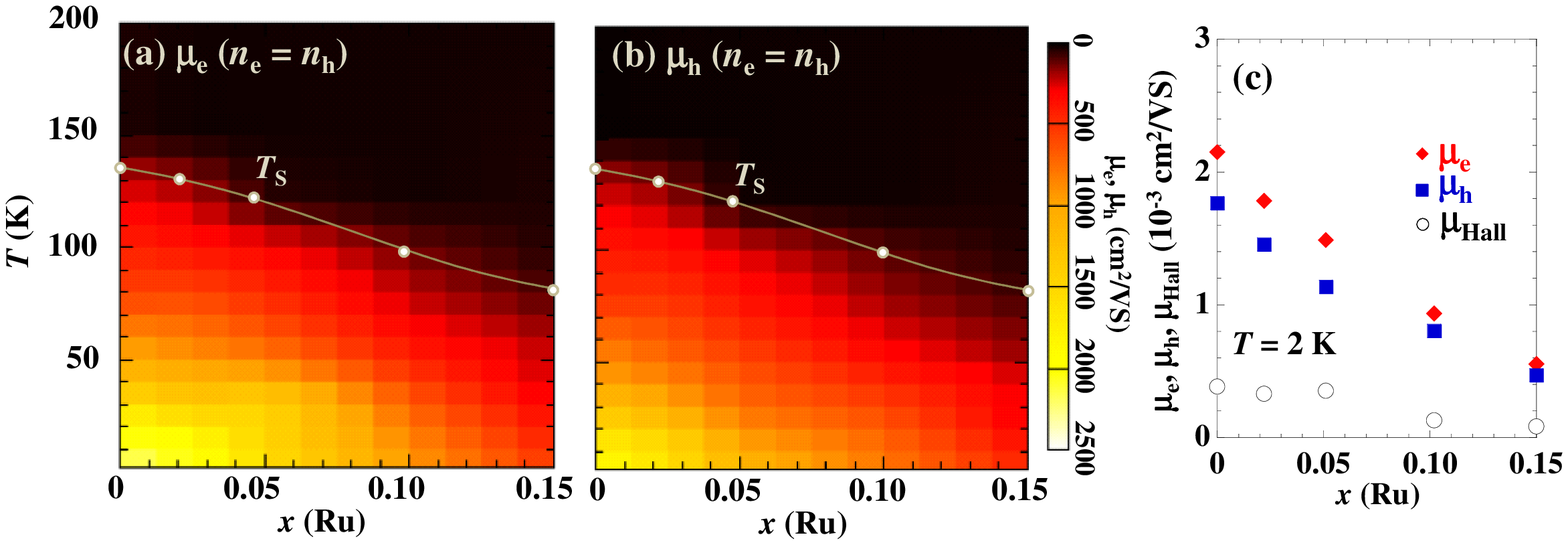}
\caption{(color online) Color contour maps of mobility for electron ($\mu$$_{\rm e}$) (a) and hole ($\nu$$_{\rm h}$) (b) in $x$ - $T$ phase diagram estimated from two-carrier-type semiclassical approximation.(c)  $x$ dependence of $\mu$$_{\rm e}$, $\mu$$_{\rm h}$, and the Hall mobility ($\mu$$_{\rm Hall}$) at 2 K.}
\end{figure*}

For estimating transport paramaters using the above three equations, an additional constraint is necessary.
It has been reported by angle resolved photo emission spectroscopy (ARPES) that the Luttinger volumes of both electron and hole Fermi surfaces are comparable with each other in Ba(FeAs)$_2$ \cite{Shen}, and that Ru substitution will not change the Fermi surface significantly \cite{Kam}.
Therefore, one can approximate the system as a semimetal, that is, $n_{e}$ = $n_{h}$ is applicable.
We systematically solved $\rho$(0), $R_{\rm M}$ and $\rho$$_{yx}$ by using equations (1) - (3) under the condition of $n_{e}$ = $n_{h}$, where $R_{\rm M}$ and $\rho$$_{yx}$ are quadratic and linear against $B$, respectively.

Figures 2 (a) and (b) show the color contour maps of $\mu$$_{\rm e}$ and $\mu$$_{\rm h}$ in the $T$ - $x$ phase diagram.
Both $\mu$$_{\rm e}$ and $\mu$$_{\rm h}$ increased with a decrease in temperature in the whole $x$ region below the structure and magnetic transition temperature ($T_{\rm s}$) \cite{Tanabe}.
The enhancement in $\mu$$_{\rm e}$ below $T_{\rm s}$ was larger than that of $\mu$$_{\rm h}$.
At low temperatures, $\mu$$_{\rm e}$ and $\mu$$_{\rm h}$ decreased as a function of $x$.
The $x$ dependence of mobility estimated from $R_{\rm H}$ ($\mu$$_{\rm H}$ = $R_{\rm Hall}$/$\rho$(0)) is shown in Fig. 2 (c) together with $\mu$$_{\rm e}$ and $\mu$$_{\rm h}$ at 2 K.
The value of $\mu$$_{\rm Hall}$ was significantly smaller than $\mu$$_{\rm e}$ and $\mu$$_{\rm h}$.
In a multi carrier system, the contribution of electron and holes on $\rho_{yx}$ compensates each other.
Therefore, the significantly smaller values of $\mu$$_{\rm Hall}$ than those of $\mu$$_{\rm e}$ and $\mu$$_{\rm h}$ may be indicative of the fact that electrons and holes contribute comparably to the electrical conductivity in Ba(Fe$_{1-x}$Ru$_x$As)$_2$.

It appears that the enhancement in mobility below $T_{\rm s}$ is related to the occurrence of the Dirac cone, because the Dirac cone states form under the SDW order states.
In the case of the Dirac cone type linear electronic dispersion, the cyclotron effective mass ($m^{\ast}$), being proportional to the inverse of mobility (1/$\mu$), is described in Eq. 4 by Fermi energy ($E_{\rm F}$), Fermi velocity ($v_{\rm F}$) and the carrier numbers of Dirac cone ($n$$_{\rm MR}$) \cite{Castro}.
\begin{eqnarray}
 m^{\ast} &=& \displaystyle{E_{\rm F}/v_{\rm F}^2 = (\pi n_{\rm MR})^2/v_{\rm F}} \label{4}\\
 B^{\ast} &=& \displaystyle{(1/2{\rm e}{\rm \hbar}v_{\rm F}^2)(E_{\rm F}+k_{\rm B}T)^2} \label{5}
\end{eqnarray}
The experimental observation of 1/$\mu$ $\propto$ $n$$_{\rm MR}$$^{1/2}$, therefore, indicates low-dissipative transport of Dirac fermions, independent of impurity scattering.
In a multiband system, the electrical transport is represented by mobilities and carrier numbers of all pockets (See Appendix).
If we approximate both magnetoresistance and the Hall coefficient in the n-carrier system using the two-carrier-type semiclassical model, transport parameters in each pocket in the n-carrier system can be merged into $n_{e}$, $n_{h}$, $\mu_{e}$, and $\mu_{h}$ in the two carrier model.
Therefore, we cannot distinguish electrons and holes in the two carrier model.
Here we thus employed the averaged mobility ($\mu_{\rm ave}$ = ($\mu$$_{\rm e}$ + $\mu$$_{\rm h}$)/2) to evaluate the dependence of mobility on the carrier number of the Dirac cones.
In order to estimate the carrier number of the Dirac cones, we employed a theoretical model for the analysis of magnetoresistance in the quantum limit of a Dirac cone \cite{Abrikosov}, because the carrier number of the Dirac cones cannot be estimated directly in the Ba(FeAs)$_2$ system by employing the two-carrier-type semiclassical approximation.
For Ba(Fe$_{1-x}$Ru$_x$As)$_2$, the evolution of linear magnetoresistance as a function of $B$ was observed under the SDW ordering state.
This is consistent with the quantum limit of the Dirac cone states \cite{Tanabe}.
Considering both the splitting of Landau levels and the thermal excitation energy, the onset $B$ ($B$$^{\ast}$) of the quantum limit of a Dirac cone  can be described by Eq. 5 \cite{Abrikosov}.
Employing the estimated $E_{\rm F}$ and $v_{\rm F}$ from Eq. 5 as shown in Fig. 3 (a) \cite{Tanabe}, we can estimate $n$$_{\rm MR}$ and $m^{\ast}$ using Eq. 4.

For Ba(FeAs)$_2$, it has been found from the first principles band calculations that four pockets including two Dirac cones with different carrier numbers take place in the SDW order state \cite{Shimojima, Yin}.
Experimentally, the Shubnikov-de Haas oscillation (ShdH) has shown three pockets, however the smallest Dirac cone cannot be observed \cite{Terashima}, while ARPES \cite{Richard}, the de Haas-van Alphen effect \cite{Harison} and the complex term in THz optical conductivity \cite{Imai} indicate the existence of another small Dirac cone.
Since the $E_{\rm F}$ estimated from the magnetoresistance is close to that of the smallest Dirac cone for Ba(FeAs)$_2$ \cite{Richard, Imai}, $n$$_{\rm MR}$ can reasonably be assigned to the smallest Dirac cone.

\begin{figure}[h]
\includegraphics[width=1.0\linewidth]{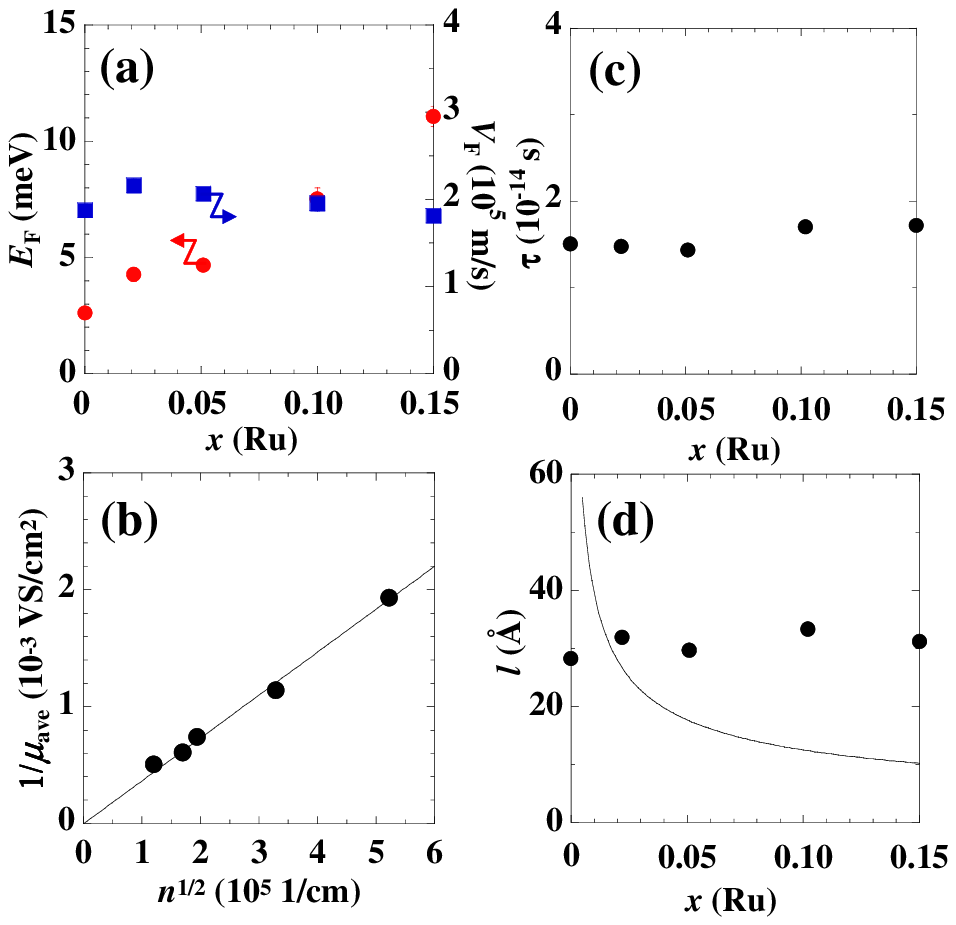}
\caption{(color online) (a)$x$ dependence of Fermi energy ($E_{\rm F}$) and Fermi Velocity ($v_{\rm F}$) of Dirac cone \cite{Tanabe} for Ba(Fe$_{1-x}$Ru$_x$As)$_2$. (b) Carrier numbers of Dirac cone ($n$$_{\rm MR}$) dependence of inverse of averaged mobility ($\mu_{\rm ave}$ = ($\mu$$_{\rm e}$ + $\mu$$_{\rm h}$)/2) for Ba(Fe$_{1-x}$Ru$_x$As)$_2$. (c), (d) $x$ dependence of relaxiation time and mean free pass for Ba(Fe$_{1-x}$Ru$_x$As)$_2$ at 2 K. A solid line shows the calculated distance between Ru impurities.}
\end{figure}

Figure 3 (b) shows the dependence of 1/$\mu$$_{\rm ave}$ on $n$$_{\rm MR}$$^{1/2}$.
The value of 1/$\mu_{\rm ave}$ developed linearly against $n$$_{\rm MR}$$^{1/2}$.
Given the continuous increase in carrier number for the two Dirac cones with different carrier numbers, both values of $m^{\ast}$ linearly evolve independently with each other as a function of square root of the carrier number of the individual Dirac cones.
Therefore, even if both Dirac cones have a significant contribution on $\mu$$_{\rm ave}$, 1/$\mu$$_{\rm ave}$ will show a linear evolution as a function of $n$$_{\rm MR}$$^{1/2}$.
Consequently, the present observations of 1/$\mu_{\rm ave}$ $\propto$ $n$$_{\rm MR}$$^{1/2}$ are consistent with the expected increase in carrier numbers with cyclotron effective mass in the Dirac cone.

In order to clarify the effect of impurity scattering on the Dirac cones, estimation of the mean free path ($l$) is important.
Employing both $E_{\rm F}$ and $v_{\rm F}$, which are associated with the smallest Dirac cone as shown in Fig. 3 (a), $m^{\ast}$ can be calculated from Eq. 4.
We accordingly estimated both the relaxation time ($\tau$) and $l$ using $\mu$$_{\rm ave}$, $m^{\ast}$ and $v_{\rm F}$.
Note that both $\tau$ and $l$ for the smallest Dirac cone may be underestimated because $\mu$$_{\rm ave}$ is also influenced by other larger pockets.
Figure 3 (c) and (d) show the dependencies of both $\tau$ and $l$ as a function of $x$.
Both $\tau$ and $l$ are almost constant with $x$, and $l$ is larger than the average distance between Ru impurities above $x$ = 0.05.
Since $l$ is underestimated for the smallest Dirac cone, the present estimation indicates that $l$ is not reduced by Ru substitution.
Therefore, the observed almost constant value of $l$, independent of the average distance between Ru impurities, is consistent with the expected large suppression of backward scattering of Dirac fermions.

\section{Summary}
We studied electronic transport of Dirac fermions in iron pnictide materials Ba(Fe$_{1-x}$Ru$_x$As)$_2$ using rapidly quenched single crystals.
The electronic transport of parabolic bands was suppressed by scattering due to the crystal lattice distortion as well as the Ru impurity, while that of the Dirac cone is not significantly influenced.
From the analyses of magnetoresistance and Hall coefficient using the two-carrier-type semiclassical approximation, it was found that the inverse of the average mobility, being proportional to the cyclotron effective mass, developed as square root of carrier number of the smallest Dirac cone.
This can be understood in terms of the intrinsic low-dissipative carrier transport of Dirac fermions.
The estimated mean free path, using the effective mass of the smallest Dirac cone \cite{Tanabe} and the average mobility obtained from the semiclassical analysis, was constant for $x$ and larger than the average distance between Ru impurities. 
This is consistent with large suppression of backward scattering in Dirac fermions.
The present results have clearly demonstrated that the low dissipative carrier transport as a consequence of the nature of the berry's phase of the Dirac cone is realized in Ba(Fe$_{1-x}$Ru$_x$As)$_2$ under the SDW ordering state. 
This suggests that a possible enhancement of the phase coherence of Cooper pairs via the Dirac cone high mobility state plays an important role for missing in temperature fluctuations in the pnictide family.

\section*{Acknowledgements}
The authors are grateful to T. Tohyama for his useful comments.
The research was partially supported by Scientific Research on Priority Areas of New Materials Science using Regulated Nano Spaces, the Ministry of Education, Science, Sports and Culture, Grant in Aid for Science, and Technology of Japan and Grant-in-Aid for Young Scientists (B) (23740251).
The work was partly supported by the approval of the Japan Synchrotron Radiation Research Institute (JASRI).

\appendix*
\section{Magnetoresistance and Hall coefficient for 4-carrier-type semiclassical approximation}

It was theoretically predicted by the band calculations employing the local density approximation and the density functional theory that the Fermi surface of Ba(FeAs)$_2$ is composed of 4 pockets including 2 Dirac cones with different carrier numbers in the SDW ordering state \cite{Shimojima, Yin}.
In this case, both magnetoresistance and Hall coefficients can be described using mobilities and carrier numbers of four pockets.
Here, we showed equations of magnetoresistance ($R_{\rm M}$) and Hall resistivity ($\rho_{yx}$) derived from the 4-carrier-type semiclassical approximation in the low-B limit.

A semiclassical-type conductivity tensor of a carrier can be described in eq. (A.1) \cite{AM}.

\begin{eqnarray}
 \sigma_i &=& \frac{en_i\mu_i}{1+\mu_i^2B^2}
\left[ 
\begin{array}{cc}
1 & \mu_iB \\
-\mu_iB & 1 \\
\end{array} 
\right]
 \label{1}
\end{eqnarray}

In the 4-carrier-type expression, the resistivity tensor can be obtained using the inverse of sum of 4 conductivity tensors.

\begin{eqnarray}
 \rho &=& (\sum^4_{i=1}\sigma_i)^{-1} = 
\left[ 
\begin{array}{cc}
\rho_{xx} & -\rho_{yx} \\
\rho_{yx} & \rho_{xx} \\
\end{array} 
\right]
\label{2}
\end{eqnarray}

In the low-$B$ limit both $R_{\rm M}$ and $\rho_{yx}$ can be approximated as shown in eqs. (A.3) and (A.4).


\begin{eqnarray}
R_{\rm M} &=& \frac{(\alpha + \beta + \gamma)B^2}{\delta} \\
\rho_{yx} &=& \frac{\epsilon B}{e\delta},
\end{eqnarray}

where
\begin{eqnarray*}
\alpha &=& ((\sqrt{\mathstrut \mu_1 \mu_4^3 n_1} - \sqrt{\mathstrut \mu_1^3 \mu_4 n_1})^2 + (\sqrt{\mathstrut \mu_2 \mu_4^3 n_2} - \sqrt{\mathstrut \mu_2^3 \mu_4 n_2})^2 \\
&&+ (\sqrt{\mathstrut \mu_3 \mu_4^3 n_3} - \sqrt{\mathstrut \mu_3^3 \mu_4 n_3})^2)n_4 \\
\beta &=& ((\sqrt{\mathstrut \mu_1 \mu_3^3 n_1} - \sqrt{\mathstrut \mu_1^3 \mu_3 n_1})^2 \\
&&+ (\sqrt{\mathstrut \mu_2 \mu_3^3 n_2} - \sqrt{\mathstrut \mu_2^3 \mu_3 n_2})^2)n_3 \\
\gamma &=& (\sqrt{\mathstrut \mu_1 \mu_2^3 n_1} - \sqrt{\mathstrut \mu_1^3 \mu_2 n_1})^2 n_2 \\
\delta &=& (\mu_1 n_1 + \mu_2 n_2 + \mu_3 n_3 + \mu_4 n_4)^2 \\
\epsilon &=& \mu_1^2 n_1 + \mu_2^2 n_2 + \mu_3^2 n_3 + \mu_4^2 n_4.
\end{eqnarray*}



\end{document}